**Predicting the long time dynamic heterogeneity in a supercooled liquid on the basis of short time heterogeneities.**

Asaph Widmer-Cooper and Peter Harrowell

School of Chemistry, University of Sydney, Sydney New South Wales, 2006, Australia

**Abstract**

We report that the local Debye-Waller factor in a simulated 2D glass-forming mixture exhibits significant spatial heterogeneities and that these short time fluctuations provide an excellent predictor of the spatial distribution of the long time dynamic propensities [Phys.Rev.Lett. 93, 135701 (2004)]. In contrast, the potential energy per particle of the inherent structure does not correlate well with the spatially distributed dynamics.

In this paper we address the problem of establishing the connection between a supercooled liquid configuration and the subsequent spatial distribution of particle dynamics. As discussed in recent publications [1,2], the aspect of the dynamic heterogeneity of a supercooled liquid that derives from the initial configuration is described by the dynamic propensity. The propensity of a particle, which we will define in detail below, is directly associated with the *probability* of a particle in a configuration undergoing a substantial displacement within a time interval, as distinct from how far it is actually observed to move in a single trajectory. This propensity for motion is the starting point for models of glass relaxation such as the facilitated spin models [3] and the cooperative lattice gas models [4]. Each of these models is defined by a set of rules in which the probability for movement is determined by the instantaneous configuration. In contrast, most molecular models of glass formers are defined by a Hamiltonian and structural constraints. In these models, uncovering the relationship between particle configurations and the probability of particle motion represents a major challenge and is the focus of this paper. Our ultimate goal is to predict the spatial pattern of dynamic propensity from a given configuration. *Prediction*, we shall see, requires more than simply the demonstration of a correlation between averages. In this paper we shall first examine how the correlation between dynamics and the energy of the local potential minimum (the inherent structure or IS) fails to provide a useful dynamic prediction. We then report on the striking success of the short time dynamics of each particle in predicting the spatial structure of dynamics over long times.

We have chosen a well-characterised model of a two-dimensional glass-forming alloy.[5] The equimolar mixture consists of particles interacting via a purely repulsive potential of the form $u_{ab}(r) = \varepsilon(\sigma_{ab}/r)^{12}$ , where $\sigma_{12} = 1.2 \times \sigma_{11}$ and $\sigma_{22} = 1.4 \times \sigma_{11}$. All units quoted will be reduced so that $\sigma_{11} = \varepsilon = m = 1.0$ where $m$ is the mass of both types of particle. Specifically, the reduced unit of time is $\tau = \sigma_{11}(m/\varepsilon)^{\frac{1}{2}}$. Details of the simulation have been described previously [1,2,5]. Following refs. [1,2], the dynamic propensity of particle i is defined as $< (\vec{r}_i(0) - \vec{r}_i(t))^2 >$, where the average is taken over the ensemble of $N$-particle trajectories, all starting from the same configuration but with momenta assigned randomly from the appropriate Maxwell-Boltzmann distribution. (We note that Doliwa and Heuer [6] have used multiple trajectories from a single configuration at *different* temperatures to establish the Arrhenius character of transitions between metabasins.) The time interval needs to be chosen to permit the observation of dynamic heterogeneities. Here we have chosen the interval to be 1.5 x $\tau_e$ where $\tau_e$, the structural relaxation time, is defined as the time at which the intermediate incoherent scattering function, measured at the wave vector of the first peak in the structure factor, has decayed to 1/e. This ensemble mean squared displacement does not correspond to the actual squared displacement of the particle in any particular run but, rather, reflects the particle's propensity for displacement. The propensity map for a configuration taken from an equilibrium distribution at T = 0.4 is shown in Figure 1. We have used a contour plot, using a Shepard interpolation [7], in order to aid in visualizing the dynamic structure. The heterogeneous structure is evident in both the localised sites of high propensity and the extended domains of low propensity.

What aspect of the initial configuration is responsible for the spatial distribution of propensity? We shall examine the correlation between the spatial distributions of particle potential energy in the initial IS and particle propensity. Doliwa and Heuer [6,8] and la Nave and Sciortino [9] have reported correlations between the dynamics of small systems (60-120 particles) and the inherent structure (IS) energy. These calculations did not look at whether the correlation extends to the spatial distribution of the two quantities, the issue we will address here. Each particle is assigned a potential energy equal to the sum over neighbour interactions (neighbours being defined by a cutoff distance equal to the position of the first minimum in the appropriate pair distribution function) in the IS. In Figure 2 we plot the spatial distribution of this energy for the same configuration as used in Figure 1. From inspection, we note that the spatial heterogeneities of the IS particle energies involve considerably shorter length scales than those of the propensity. This is confirmed by comparing the number of clusters formed by particles in the top 10% of propensity and potential energy, as shown in Figure 3 for 10 configurations (see the Figure caption for the definition of a cluster). The apparent contradiction between these results and the previous reports [6,8,9] underscores the difficulty of interpreting correlations. Elsewhere [10] we have shown that correlations between *average* values of two quantities does not necessarily mean that a microscopic, and hence causal, correlation exists. The typically large standard deviations of these quantities are evidence of this.

We have recently shown [10] that the local free volume does no better than the local potential energy as a predictor of the propensity. The local composition has also proven to be of limited predictive use. In the 2D mixture, we find roughly 23 local compositional environments, most exhibiting propensities that span the entire liquid range [2]. In the face of this persistent failure of local structural measures to predict the spatial distribution of propensity we have reconsidered the question of what ultimately determines a particle's ability to move. The ability to move is associated with the degree to which particles are constrained by their surroundings. The proposal that potential energy or free volume would correlate with local mobility rests on the expectation that these local scalar measures capture an essential aspect of this constraint. Having found this not to be the case, we now consider the nature of local constraints explicitly.

Thorpe [11], in studies of network glass formers, has shown how the lack of full constraint is manifest as floppy modes, an observation that clearly offers a connection between a configuration and its dynamic heterogeneity. This constraint counting has not been applied to glasses stabilized by dense packing (as opposed to directional bonds) because of the unsolved problem of identifying local constraints in the former case. We shall sidestep this problem by looking directly for floppy modes, rather than trying to guess how to quantify the constraints responsible. We have collected statistics of individual particle motion by running many short trajectories from the same configuration, again randomizing the momenta. The duration of the run is short, $10\tau$, which, at T = 0.4, lies in the middle of the plateau in the log-log plot of $<\Delta r^2(t)>$ vs time and is a characteristic time for $\beta$ relaxation. We shall refer to the resulting variance of the $i$th particle position as the local Debye-Waller (DW) factor for particle $i$. From comparison of the IS configurations at either end of the $10\tau$ intervals we find that this time roughly correspond to the first 'escape' from the initial IS, involving a small localized reorganization of particles.

As shown in Figure 3, the local DW factors result in a relatively small number of clusters - evidence of a substantial and nontrivial heterogeneity. Furthermore, the number of these clusters is quite similar to that produced by the propensities. How well do the local DW factors predict the spatial distribution of the propensity? To use the DW factors as a predictor of propensity we shall require them to meet two criteria: they must exceed a Lindemann-like threshold of 0.035 [12], and the particles must be in a cluster of three or more. We find that the selected particles provide an excellent prediction of the spatial variation of propensity. In Figure 4 we compare the prediction of high propensity using the particle DW factors with the propensity maps for six independent configurations. We find that the prediction of the high propensity regions from the local DW factors is very good and very few points lie in regions of low propensity (in contrast, for example, with the predictions for high propensity based on the potential energy maps). Our data supports the proposition that the high DW regions represent the precursors to the long time motion and that the subsequent propagation of the consequences of these 'seed' motions is not readily accessible from the initial configuration, hence the coarse grained character of the DW factors' predictive success.

We have shown that in a supercooled liquid the variance of the particle displacements, averaged over an iso-configurational ensemble, is heterogeneously distributed. This result is consistent with the observation of heterogeneous displacements found previously below $T_g$ [13]. More importantly, we have shown that the spatial distribution of the local DW factor, a measure of short time motion, is very similar to that of the dynamic propensity, which reflects dynamics over time scales two orders of magnitude longer. Invoking criteria inspired by a heterogeneous extension of the Lindemann melting criterion for amorphous materials [12], we find that the DW factors provide an excellent predictor of the spatial distribution of the high propensity domains in each configuration studied. This success is the more striking when compared with the absence of any strong correlation between propensity and quantities like the local energy or free volume. We conclude that the initial configuration determines the local DW distribution (corresponding to the β processes), which in turn is the precursor to the subsequent dynamic propensity (characteristic of the α process). These results extend the growing experimental [14, 15] and theoretical [16] evidence for correlations between high and low frequency response to the spatial heterogeneities of the two processes. The demonstration that short runs predict most of the dynamic heterogeneity that arises due to the initial configuration represents a significant speed up (by at least a factor of 100) of the task of establishing the dynamic propensity.

Given the subtlety of the collective mechanical constraints probed by the short time dynamics, it is very unlikely that any measure of the initial configuration will provide a better prediction of the dynamic propensity than that shown in Figure 4. Subsequent answers may improve the algorithmic efficiency in mapping between configuration and the selected DW map but it is unlikely that they will improve upon the quality of the answer. If this proposal is accepted then one has, in this work, a sense of the limits one should expect in the answer to the core problem of the glass transition, i.e. the causal connection between structure and dynamics. Wolfram [17] has pointed out that there are phenomena in complex systems that are irreducible, in the sense that the future behavior cannot be obtained by an algorithm more efficient than the solution of the equations of motion. Recently Israeli et al [18] have qualified this observation by noting that prediction is possible for suitably coarse-grained versions of the outcome. Our results certainly support the idea that judicious coarse graining of the structure-dynamic problem is an important part of obtaining a satisfactory solution. Work continues on the complete elimination of dynamics in the prediction of the propensity from the structure.

**Acknowledgements** We gratefully acknowledge support of the Discovery program of the Australian Research Council.

**Figure Captions**

Figure 1. A map of the dynamic propensity for a configuration taken from an equilibrated system at T = 0.4. 100 runs were averaged over to create the map.

Figure 2. A map of the potential energy per particle for the inherent structure of the same configuration used in Figure 1.

Figure 3. A cluster analysis of the spatial distribution of propensity, Debye-Waller factors, and potential energy per particle for 10 independent configurations at T=0.4. To measure the spatial heterogeneity of a given property $S$ we 'tag' the 10% of particles with the largest values of $S$. We then assign each tagged particle to a cluster if it is a neighbour to another tagged particle already in that cluster (neighbours being defined by a cutoff distance equal to the position of the first minimum in the appropriate pair distribution function). When all the tagged particles have been assigned to a cluster we count the number of clusters. 'Random' refers to the number of clusters generated by a random distribution of 102 particles.

Figure 4. A comparison of the predictions of high propensity (filled circles) based on the local DW data as described in the text with the actual propensity distributions for 6 independent configurations taken from an equilibrated system at T = 0.4. The colour scale is the same as that used in Figure 1.

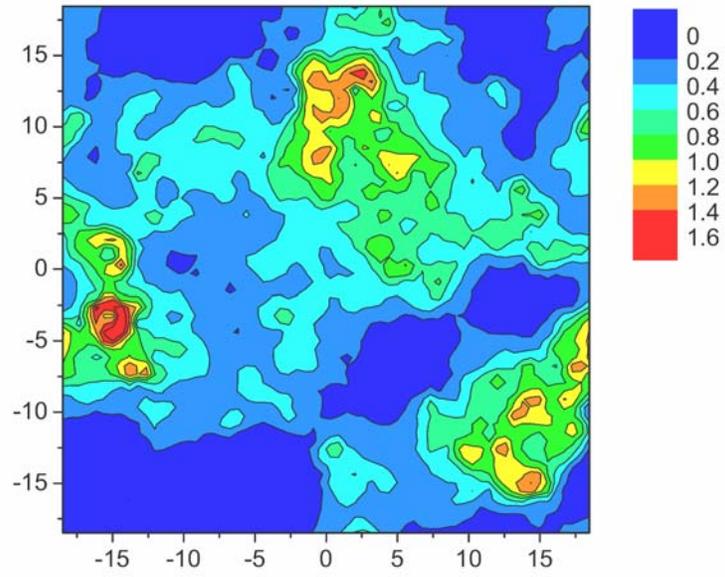

Figure 1

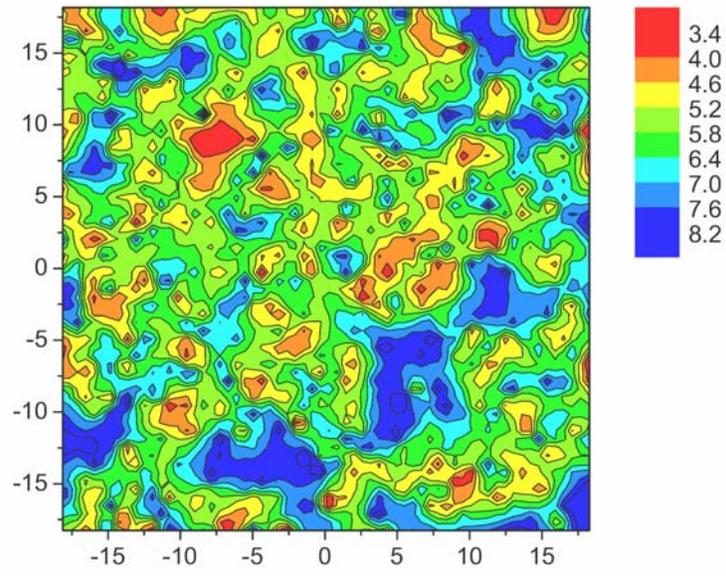

Figure 2.

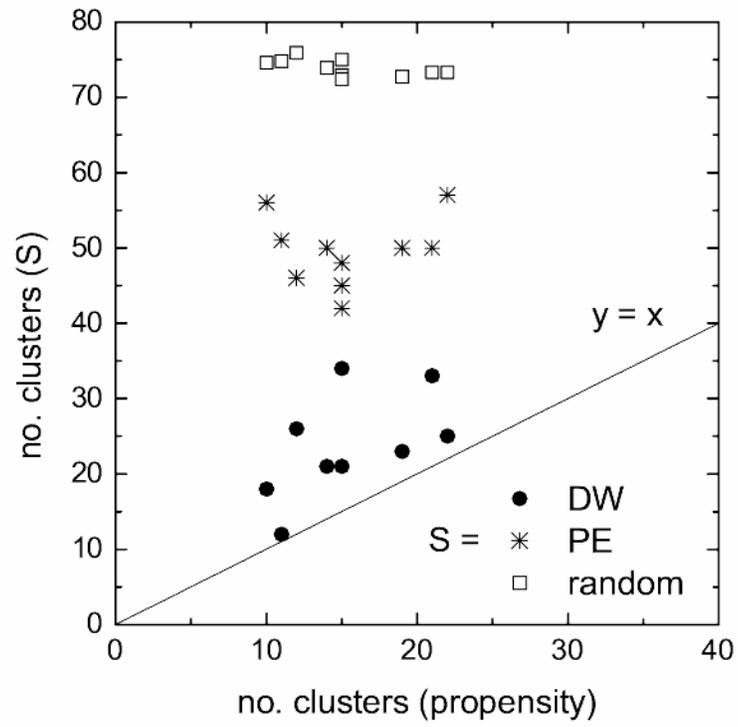

Figure 3

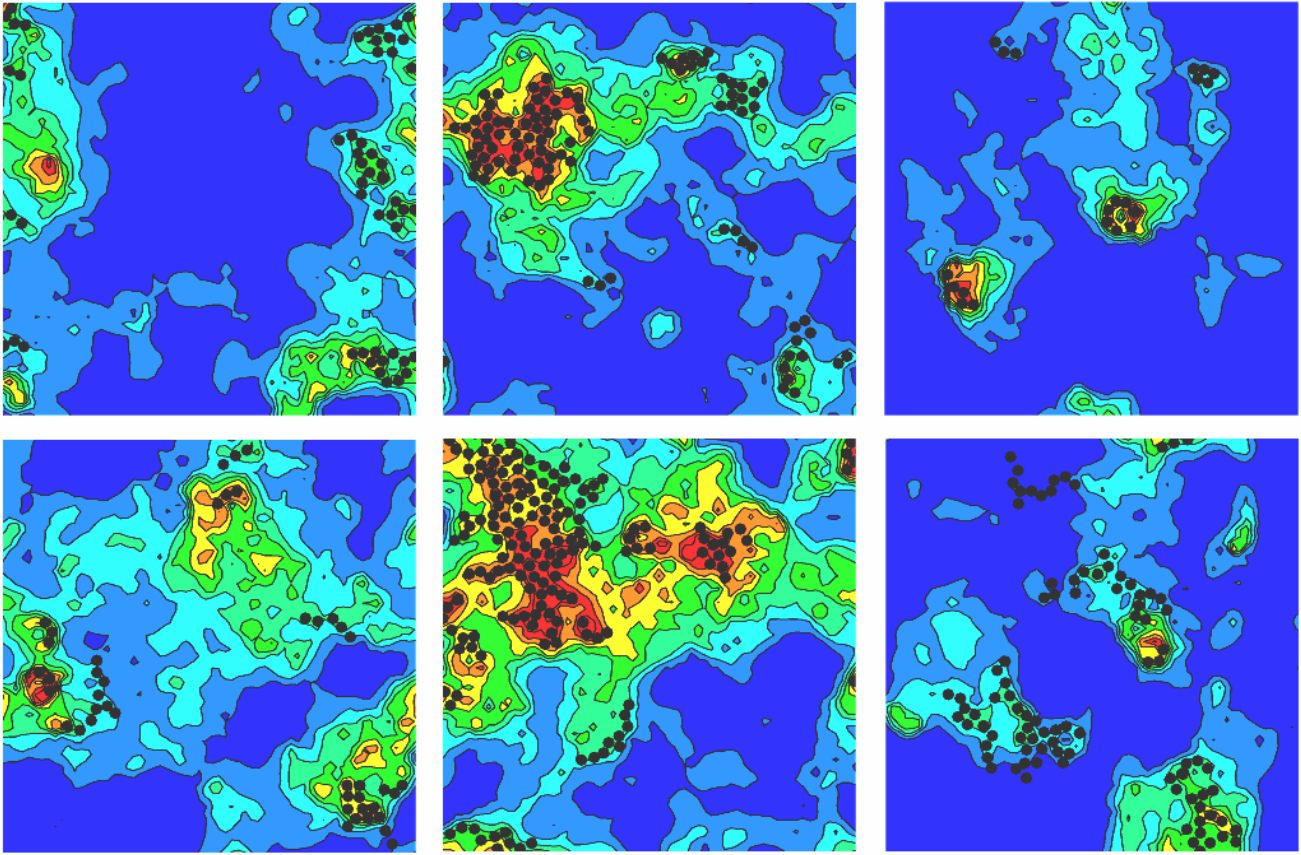

Figure 4.